\documentclass[ammsymb,pre,aps,twocolumn,superscriptaddress]{revtex4}
\usepackage{graphicx}

\def \beq{\begin{equation}}
\def \eeq{\end{equation}}
\def \beqar{\begin{eqnarray}}
\def \eeqar{\end{eqnarray}}
\input{tcilatex}
\begin{document}

\title{Does Phylogenetic Proximity Explain Nestedness in Mutualistic
Ecosystems?}
\author{R.P.J. Perazzo}
\affiliation{Departamento de Investigaci\'{o}n y Desarrollo, Instituto Tecnol\'{o}gico de
Buenos Aires \\
Avenida E. Madero 399, Buenos Aires, Argentina}
\author{Laura Hern\'{a}ndez}
\affiliation{Laboratoire de Physique Th\'eorique et Mod\'elisation; \\
UMR CNRS, Universit\'e de Cergy-Pontoise, \\
2 Avenue Adolphe Chauvin, 95302, Cergy-Pontoise Cedex France}
\author{Horacio Ceva}
\affiliation{Departamento de F{\'{\i}}sica, Comisi{\'o}n Nacional de Energ{\'\i }a At{\'o}%
mica,\\
Avenida del Libertador 8250, 1429 Buenos Aires, Argentina}
\author{Enrique Burgos}
\affiliation{Departamento de F{\'{\i}}sica, Comisi{\'o}n Nacional de Energ{\'\i }a At{\'o}%
mica,\\
Avenida del Libertador 8250, 1429 Buenos Aires, Argentina}
\affiliation{Consejo Nacional de Investigaciones Cient\'{\i}ficas y T\'{e}cnicas,\\
Avenida Rivadavia 1917, C1033AAJ, Buenos Aires, Argentina}
\author{Jos\'{e} Ignacio Alvarez-Hamelin}
\affiliation{Departamento de Investigaci\'{o}n y Desarrollo, Instituto Tecnol\'{o}gico de
Buenos Aires \\
Avenida E. Madero 399, Buenos Aires, Argentina}
\affiliation{Consejo Nacional de Investigaciones Cient\'{\i}ficas y T\'{e}cnicas,\\
Avenida Rivadavia 1917, C1033AAJ, Buenos Aires, Argentina}

\begin{abstract}
ABSTRACT

We investigate how the pattern of contacts between species in mutualistic
ecosystems is affected by the phylogenetic proximity between the species of
each guild. We develop a dynamical model geared to establish the role of 
such proximity in the emergence of a nested pattern of contacts. We also define  
a parameter that provides a direct measure of the influence of phylogenetic      
proximity in a given pattern of contacts. We conclude that although phylogenetic 
proximity is compatible with
nestedness it can not be claimed to be a cause of it. We find that nestedness
can instead be attributed to a general rule by which species tend to
hold contacts with counterparts that already have a large
number of contacts. If the phylogenetic structure of both guilds is brought
into the analysis, this rule is equivalent to maximize the phylogenetic
diversity of the mutualistic counterparts of species of either guild.
\end{abstract}


\maketitle 

keywords: Nested networks; Mutualistic communities; Phylogenetic proximity;
Ultrametricity.

Mathematics Subject Classification  65C20  68U20  90C27  90C35  92B10

Corresponding author: H. Ceva; ceva@cnea.gov.ar;  FAX: 54-11-6772-7121

\section{Introduction}

A sustainable management of ecosystems as well as a proper assessment of the
impact of human activity on them can only be achieved with a proper
understanding of the pattern of the interactions between the species. We are
here interested in the case of mutualistic systems. These usually involve
groups of animals and plants, helping each other to fulfill essential
biological functions such as feeding or reproduction. This is the case of
systems in which animals feed from fruits while dispersing the seeds (%
\textit{seed dispersal networks}) or those where insects feed from the
nectar of flowers while helping the plant in the pollination process (%
\textit{pollination networks}).

The structure of such systems is described by means of an adjacency matrix
whose elements represent the absence or presence of an interaction between
the plant and animal species. In mutualistic networks this matrix strongly
indicates that the mutualist ecosystems are not a random collection of
interacting species, but that they display instead, a high degree of
internal organization. A pervading feature that has been observed is that
the adjacency matrix has a nested pattern of interactions, in which both
generalists (species holding many interactions) and specialists (holding few
interactions) tend to interact with generalists whereas
specialist-to-specialist interactions are infrequent \cite{nosotros2}. In
other words, if species are ordered by decreasing number of contacts, then
the contacts of a given species constitute a subset of the contacts of the
preceeding species in the list \cite{muchas}.


The nested structure of mutualistic networks has been attributed to a number
of different causes and the controversy about the ultimate reasons that make
this pattern so frequently observed still remains. It is fairly obvious that
a detailed explanation of the interaction behavior of individual species can
be of little help to understand such a generalized pattern that is found
across ecological systems of very different sizes and types, and involving
plants of different nature and animals that range from insects to birds.

In Ref.\cite{nosotros2} it is claimed that such order may offer some
advantage for the robustness of the whole system thus suggesting that
systems that are currently observed are those that have survived less
disturbed thanks to its nested structure. Other, more elaborated theories
have been proposed. In Ref.\cite{feno} nestedness has been attributed to
phenotypic affinity between species of different guilds while in Refs.\cite%
{Rezende}, \cite{Rezende2} an extensive analysis is made concluding that
phylogenetic proximity could explain the nested organization of contacts of
some cases of mutualistic systems. In Ref. \cite{comment}, on the contrary,
the modest percentage correlations found between phylogenetic relatedness
and ecological similarity, are taken to mean that phylogenetic relationships
do not have a marked effect.

It has been customary to consider that the occurrence of some positive
statistical correlation is a sign of causation for the occurrence of the
nested pattern of contacts. However, the sole fact that in a part of the
empirical observations two elements appear to be statistically correlated
should not be taken to mean that one is the cause of the other. Such
correlation may rather indicate instead that both elements are not
incompatible, i.e., that they do not mutually exclude each other or that
they stem from a third, common cause.

One example of this analysis is given by the strong positive correlation
found between the species' abundance and hence the frequency of
interactions, with the pattern of contacts of some species \cite{Vazquez and
Aizen}. It has been suggested that locally abundant species are prone to
accumulate interactions and conversely rare species are prone to lose them 
\cite{Stang}, as also suggested by neutral theories \cite{natacha}.

One alternative way to search for causal relationships is to explore the
possible dynamic consequences of some assumed interaction mechanism, thus
verifying or falsifying hypotheses concerning possible interaction
mechanisms between the species. In Refs. \cite{nosotros1}, \cite{nosotros3}
we have proposed a Self Organizing Network Model (SNM), that allows to study
the contact pattern of a system that is consistent with some hypothetical
interaction mechanism between mutualistic species.

In the present paper we develop a modification of the SNM to take into
account the effects of phylogenetic proximity in the buid up of the contact
pattern of the system. We also apply it to investigate the stability of such
pattern of a real mutualistic system. The modified algorithm of the SNM
includes the effects of phylogenetic proximity. Mathematically such
proximity is accounted for through a matrix of distances separating any two
species of each guild of the mutualistic system. The distance matrix is
directly obtained from the topology of the phylogentic tree. We aim in this
way at establishing whether such proximity can be taken to be responsible
for the emergence of a nested pattern of contacts.

\section{Methods: the numerical modelling}

\subsection{The SNM}

Mutualistic systems can be analyzed as bipartite graphs~\cite{newman}. The
interaction pattern is usually coded into a (rectangular) adjacency matrix
in which rows and columns are labeled respectively by the plant and animal
species. Its elements $K_{p,a} \in\{0,1\}$ represent respectively the
absence or presence of an interaction (contact) between the plant species $p$
and the animal species $a$. The number of contacts of each species is the
degree of the corresponding node in the bipartite graph.

Several reasons have been given to explain the pattern of interactions
between the two guilds of a mutualistic network. They have been usually
based on positive statistical significance of correlations. One way to
elucidate a possible \textit{causal} link between some hypothetical
interaction mechanisms between mutualistic species and the pattern of
contacts, is to use a dynamical model.

The basic idea behind this strategy is to verify the \textit{consistency} of
the empirically observed contact pattern, and some hypothetical interaction
rule that may favor or hamper the contact between mutualistic species. Such
interaction can be assumed to be governed by, say, phenotypic
complementarity, phylogenetic affinity, degree, or any other possibility. We
refer to such interaction mechanism as a contact preference rule (CPR) in
the sense that it is assumed that species that verify that rule tend to hold
contacts among each other.

If a dynamical model is used having an assumed CPR as an ingredient, any
lack of consistency between such CPR and the observed pattern of contacts
will show up as instabilities of the results of the model. These may easily
be detected when the observed adjacency matrix tends to be drastically
altered if species are allowed to redefine their contacts to better fulfill
some assumed CPR. On the other hand, the contact pattern and the assumed CPR
can be seen to be consistent with each other if the contact pattern only
suffers mild alterations when species redefine their contacts according to
the CPR. The redefinition of contacts in accordance with some given CPR is
accomplished with the SNM that we briefly explain below.

Within the same line of thought, it is also possible to check which is the
contact pattern that would \textit{emerge} from a random adjacency matrix if
iterated changes according to a given CPR are allowed to take place. An
emerging consistent contact pattern gives a clue of what one should expect
to observe in nature if some given CPR is the prevailing interaction
mechanism among the species of system.

From a purely theoretical point of view this set up is equivalent to
consider that the observed pattern of interactions corresponds to an \textit{%
optimal} assignment of the contacts between both guilds, with two
constraints. The first constraint is the fulfillment of an assumed CPR, and
\ the second, is the given (constant) total number of contacts between the
two mutualist guilds. This number might be considered as an indicator of the
energy invested by all the species of the ecological system in their
exchange of nourishment. In other words one may attempt to describe the
observed pattern of contacts as the result of a (combinatorial) \textit{%
optimization} problem by which contacts in the adjacency matrix are placed
in such a way as to reach an extreme of an utility function that corresponds
to an optimal fulfillment of some prevailing CPR.

An example of a dynamical model incorporating some predefined CPR is the SNM
developed in refs. \cite{nosotros2},\cite{nosotros1}. Within this model, the
mutualistic system is assimilated to a bipartite graph and the topology of
that network is established as the result of a self-organization process.
This amounts to redefine links gradually and progressively alternating
plants and animals. Say a plant is first chosen at random and one of its
contacts is redefined by spotting a mutualistic counterpart with which it is
possible a better fulfillment of the prevailing CPR. Next an animal is
chosen and the same procedure is accomplished. An iteration of these steps
provides a simple heuristic, leading to a good approximation of the optimal
assignement of contacts mentioned above.

Unlike in the preferential attachment algorithm \cite{Barabasi}, in the SNM
the topology of a \textit{non-growing} network with a fixed number of nodes
is progressively reshaped: in each iteration a connection between two nodes
of a different kind is rewired to favor a contact with the node having the
highest degree. It is worthwhile noting that in this sense, the CPR of the
SNM is \textit{local}: it doesn't take into account the whole probability
distribution, but only the degrees of the two randomly chosen species.

In the above references we show that that CPR always leads to nested
networks with degree distributions that closely resemble the ones reported
from the observation of real mutualistic systems.

It is worthwhile to point out that the self-organization process \textit{%
does not pretend to represent a real life behavior} of plants and animals of
the system. A starting random adjaceny matrix is used as an initial
condition for the dynamical process with the least possible bias. The whole
process therefore does not aim at reproducing an evolutive or adaptive
process. It rather provides a plausible mathematical tool to search for the
pattern corresponding to an optimal assignment of contacts as explained
above or to check for consistencies between some assumed CPR and the
observed data.

The SNM that we have just described has to be modified to take into
consideration the phylogenetic structure of both guilds. Therefore, it is
necessary to have a simple quantitative measure of the phylogenetic
structure of both guilds. We now describe how this is made.

\subsection{The Ultrametric Organization of Phylogenetic Trees}

The classification of species according to their similarities has been a
major endeavor since the origins of biology as a natural science. The
starting point of a classification of $N$ species along this line is to have
a symmetric $N\times N$ distance matrix with vanishing diagonal elements
providing a measure of similarities and differences between any pair of
species. Due to the central role of evolution, these classifications are
depicted by phylogenetic trees that are determined using several sources of
information.

\textit{\ }Comparative studies of phenotypic traits are also widely used.
The resemblance of species is measured through a phylogenetic signal that is
quantitatively estimated through statistical analyses \cite{Felsestein}, 
\cite{Blomberg} of the distribution of the values of different traits. These
studies may also be suplemented whenever possible with fossil records.

While the tips of the tree correspond to presently observed species, the
remaining nodes are associated to their presumed ancestors. A hierarchical
organization of all living species is therefore provided and those that
closely resemble each other are neighboring tips of the tree.

The phylogenetic classification of a group of species gather them in taxa
within taxa of an ever increasing generality. This kind of \ arrangement
gave place to pioneer taxonomic works using the concept of what now is know
as ultrametric distance. In spite of the fact that ultrametricity had being
used in biology since quite a long time (see, for instance, Ref.\cite{kimura}%
), up to that moment this was a notion exclusively used by some
mathematicians.

From a mathematical point of view, whenever this situation prevails the
objects that are classified are said to be elements of an \textit{%
ultrametric space}\cite{ultrametricos}. Based on the topology of the tree it
is possible to define the \textit{ultrametric distance} that provides a
quantitative estimation of resemblances and differences between them. In the
appendix we describe the simple procedure to extract a distance matrix
directly from the topology of the phylogenetic tree.

An ultrametric distance matrix $\ d(k,k^{\prime })$ \ constructed in this
way is not only fully consistent from the start with the results of
statistical analyses but also fully agrees with what can be expected from an
intuitive picture: small values of $d(k,k^{\prime })$ remain associated to
species that share the same branching sequence and a common evolutionary
history while large values correspond to species that have followed
different evolutionary process because they have been separated at earlier
stages. With these conventions the closest possible distance between any two
species is 1 and, if all species are at a distance 1 they belong to a 
\textit{star phylogeny}.

\subsection{The phylogenetic-SNM}

\label{resultados}

We are now in the position to define the self organization process involved
in the SNM using phylogenetic relatedness among the species of each guild.

The influence of phylogenetic proximity in the pattern of contacts can be
cast into a modified SNM by properly defining a CPR based on the ultrametric
distance between species of each guild.

In each step of this modified SNM a link of a randomly chosen species is
also redefined. However this is made in such a way as to better fulfill a
CPR by which phylogenetically close species - i.e. separated by a short
ultrametric distance - tend to have the same contacts. The basic idea is to
check either if this CPR gives rise to a nested pattern of contacts in the
adjacency matrix when starting from a random initial condition or, if an
initially nested pattern is stable under a dynamics induced by that CPR.

For each time-step of the algorithm, the following actions are performed:

\begin{itemize}
\item Two elements $K_{p,a_{1}}=1$ and $K_{p,a_{0}}=0$ are chosen at random
in the same row $p$, corresponding to connected and disconnected species
respectively.

\item Two total ultrametric distances $S_{0}$ and $S_{1}$ are calculated
between the plant $p$ and the other plants $p\prime $ which hold a contact
with $a_{0}$ and $a_{1}$ respectively.

\begin{eqnarray}
S_{0} &=&\sum_{p^{\prime }}d(p,p^{\prime })K_{p^{\prime },a_{0}} \\
S_{1} &=&\sum_{p^{\prime }}d(p,p^{\prime })K_{p^{\prime },a_{1}}
\end{eqnarray}

Notice that $S_{0}$ and $S_{1}$ are the total ultrametric distances
separating the randomly chosen plant from all other plants holding contacts
with the same animal. $S_{1}$ corresponds to the current contact and $S_{0}$
is calculated for an alternative location of the contact.

\item A swap between this two elements corresponding to the redefinition of
the link, i.e., $K_{p,a_{0}}^{\mbox{new}}\rightarrow 1$ and $K_{p,a_{1}}^{%
\mbox{new}}\rightarrow 0,$ is proposed and it will be accepted if the
following two conditions are statisfied:

\begin{itemize}
\item {(i)} neither animal species $a_{1} $ nor $a_{0}$ become extinct due
to the swapping,

\item {(ii)} $S_{1}\geq S_{0}$

In this case $a_{0}$ is a better multualistic counterpart of $p$; it belongs
to a group whose members are closer phylogenetic relatives than the original
group of contacts of $a_{1}$. Whenever the conditions (i) or (ii) are not
met, the proposed swapping is rejected.
\end{itemize}
\end{itemize}

These steps are repeated, alternatively inverting the role of rows and
columns, until the algorithm converges (no more changes are possible).

The CPR just described, which we call MIN-CPR, is certainly not the unique
way to take into account the phylogenetic distance. It is interesting to
consider the opposite rule, which we call MAX-CPR and which consists in
replacing the condition (ii) by $S_{1}\leq S_{0}$ to accept the swapping.

Since the MIN-CPR and MAX-CPR conditions are mutually exclusive we run two
independent SNM algorithms.

Within the MIN-CPR possibility, the animal counterpart that will finally be
selected for the plant $p$, is such that the set of plants $p^{\prime }$
having contact with the animal $a_{0}$ are phylogenetically closer to $p$
than those in contact with $a_{1}$. This is so because the sums $S_{0}$ and $%
S_{1}$ involve all distances between the plant $p$ (that has been selected
at random) and all other plants that hold contacts with the two animals $%
a_{0}$ and $a_{1}$. With this algorithm the configuration of contacts is
progressively dominated by phylogenetic proximity as measured by the
corresponding distance matrices. Within this strategy species of one guild
are assumed to interact in the same fashion as all other species of the same
guild belonging to their phylogenetic neighborhood.

Within the MAX-CPR possibility, the animal counterpart that will be selected
for the plant $p$ is such that all other plants $p^{\prime }$ that have some
contact with it are phylogenetically further from $p$. In this way the set
of species that share contacts with $p$ tend to have a greater phylogenetic
diversity. This is so because the pattern of contacts is progressively
dominated by greater phylogenetic distances as measured by the corresponding
distance matrices. In this alternative all species tend to be as generalist
as possible in what refers their phylogenetic grouping.

In order to check the ordering process generated by the SNM we define an 
\textit{effective distance between interacting species} as: 
\begin{equation}
D^{A,P}=\frac{1}{\langle d_{A,P}\rangle }\frac{\sum_{k,k^{\prime
}}d_{A,P}(k,k^{\prime })\tilde{W}^{A,P}_{k,k^{\prime }}}{\sum_{k,k^{\prime }}%
\tilde{W}^{A.P}_{k,k^{\prime }}}  \label{paramD}
\end{equation}%
In Eq.(\ref{paramD}) A and P respectively represent animals and plants, and $%
\tilde{W}$ represents the unweighted adjacency matrix of the projected
graphs for animals or plants that are defined as (with $K^{T}$ being the
transposed of $K$) 
\begin{eqnarray}
W_{p,p^{\prime }}^{P} &=&\sum_{a}K_{p,a}K_{a,p^{\prime }}^{T}(1-\delta
_{p,p^{\prime }})  \label{proyectado1} \\
W_{a,a^{\prime }}^{A} &=&\sum_{p}K_{a,p}^{T}K_{p,a^{\prime }}(1-\delta
_{a,a^{\prime }})  \label{proyectado2}
\end{eqnarray}%
and $\tilde{W}_{k,k^{\prime }}=1$ if $W(k,k^{\prime })\neq 0$ or 0 if $%
W(k,k^{\prime })=0,$ i.e., two species share or do not share mutualistic
counterparts; $d(k,k^{\prime })$ stands for the same ultrametric distance
matrix that has been used in the self organization algorithm. This equation
provides different results for plants or animals and should therefore be
evaluated separately for the two guilds.

The sum in the denominator of Eq.(\ref{paramD}) is just the number of terms
appearing in the numerator, therefore $D$ represents the average distance
between species of the same guild that share at least one counterpart of the
other guild. $D$ is measured in units of the average distance $\langle
d\rangle $ between \textit{all} plant (animal) species of the system, namely 
\begin{equation}
\langle d_{A,P}\rangle =\frac{\sum_{k,k^{\prime }}d_{A,P}(k,k^{\prime })}{%
N_{A,P}(N_{A,P}-1)}
\end{equation}%
where $N_{A,P}$ is the number of animal or plant species of the system.

A value $D^{A,P}\ $significantly less than unity indicates that phylogenetic
proximity is a dominant effect, because species that share at least one
mutualistic counterpart are closer than the average separation of species of
that guild. If species are ordered as the tips of a phylogenetic tree, $%
\tilde{W}_{k,k^{\prime }}$ has most of its non vanishing elements close to
the diagonal. If $D^{A,P}\geq 1$ or $D^{A,P}\simeq 1$ is instead a signature
that phylogenetic proximity is not relevant. Correspondingly $\tilde{W}%
_{k,k^{\prime }}$ tends to display non vanishing elements far from the
diagonal. The value of $D^{A,P}$ is therefore a good order parameter
characterizing the whole pattern of contacts and can also be used to check
if the the ordering process implied in the SNM converges to stable
configuration.

\begin{center}
\begin{figure}[tbp]
\includegraphics[width=9.5cm]{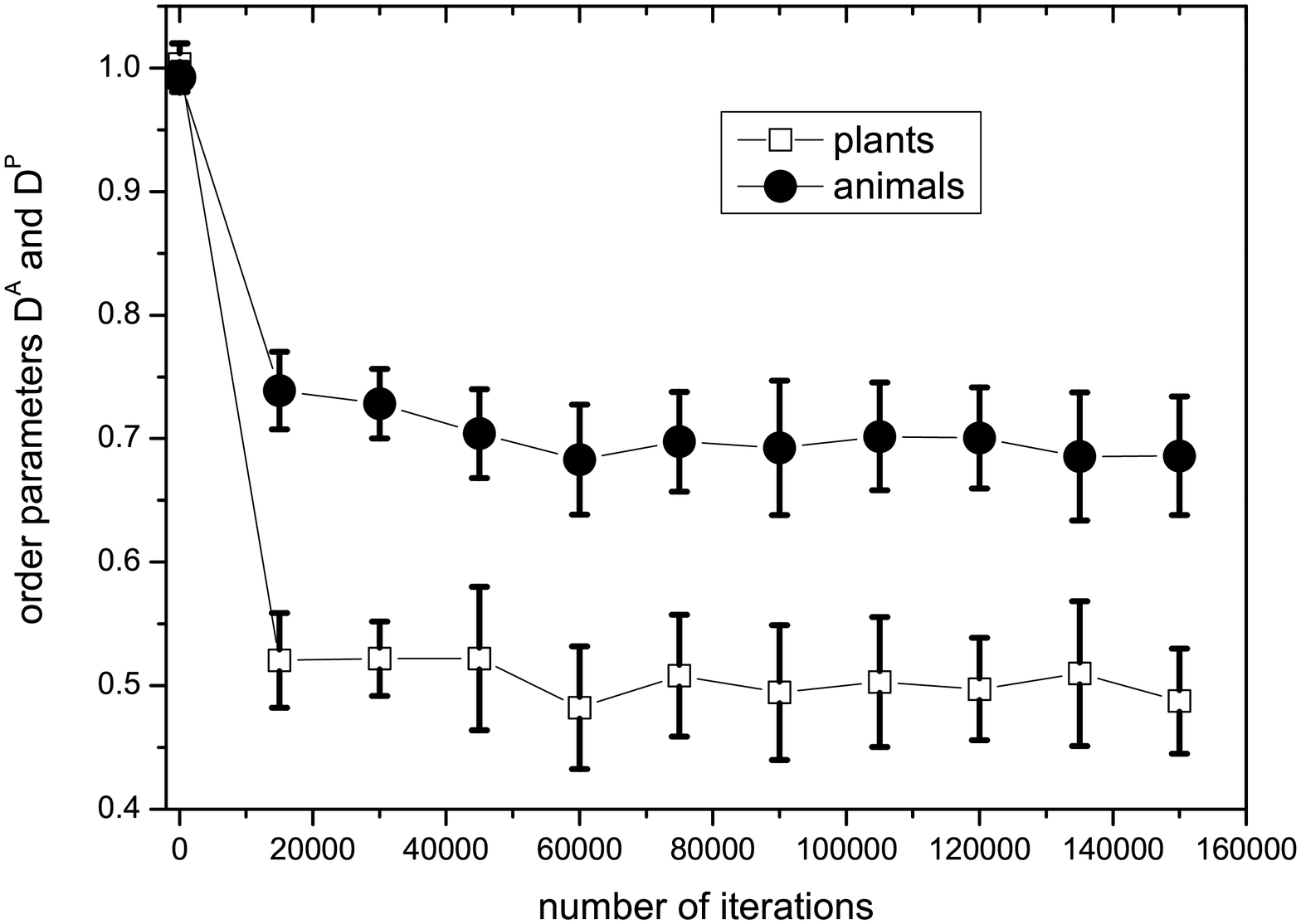}
\caption{Order parameters $D^{A}$ and $D^{P}$\ (effective distance between
species sharing the same mutualistic counterparts), as a function of the
number of iteration steps of the SNM algorithm, using the MIN-CPR
alternative. The curves are the average over several realizations with
random initial conditions, always using the same number of rows, columns,
and contacts than the system NCOR. Error bars are the corresponding standard
deviations.}
\label{parorden1}
\end{figure}

\begin{figure}[tbp]
\includegraphics[width=9.5cm]{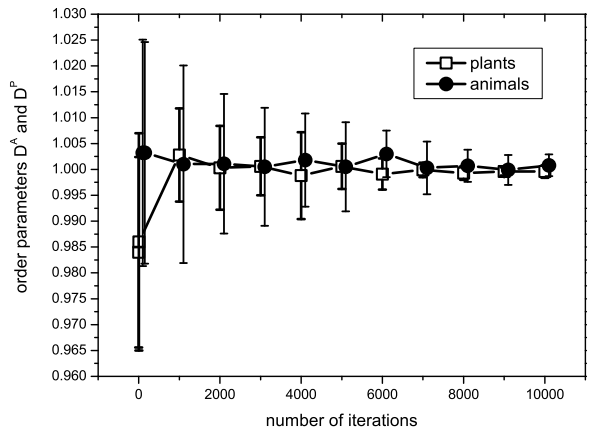}
\caption{Value of the order parameters $D^{A}$ \ and $D^{P}$ as a function
of the number of iteration steps of the SNM algorithm using the MAX-CPR
alternative. Notice that the vertical scale is strongly expanded as compared
to that of Fig.\protect\ref{parorden1}. Conventions are the same as in that
figure. Data for plants and animals have been slightly horizontally
displaced, to facilitate its observation.}
\label{parorden2}
\end{figure}
\end{center}

\section{Results}

We will concentrate our discussion on the case of the ecosystem Nava de las
Correhuelas (NCOR) as reported in Ref. \cite{Rezende}. This system has been
presented as an example where phylogenetic proximity explains the contact
pattern between mutualistic guilds and is therefore a very good case of
study.

In Figs.\ref{parorden1} and \ref{parorden2} we plot the values of the
parameters $D^{A}$ and $D^{P}$ as a function of the number of iteration
steps of the SNM for animals and plants for the two possible CPR's
considered above.

The initial conditions are always adjacency matrices of the same number of
rows and columns and with the same number of contacts than the observed NCOR
system, except for the fact that all contacts are randomly distributed. The
curves are the average over a statistically significant number of
realizations of such random initial conditions. The error bars are the
standard deviation of the results obtained over such set. These error bars
provide also a measure of the convergence of the SNM.

As the number of SNM iterations grows both CPR's produce values of $D^{A}$
and $D^{P}$ that reach asymptotic constant values. This indicates that both
CPR's suceed in driving the system to a stable and ordered pattern of
contacts. These are however different. While for the MIN-CPR case it is
found that the asymptotic value of the order parameters $D^A$ and $D^P$
stabilizes at values that are significantly smaller than unity, for the
MAX-CPR alternative it is found instead that they reach a stable value of 1.

A value $D^A \simeq {D^P} \simeq 1$ indicates that the distance between
species of the same guild that share the same mutualistic counterparts is
close to the average distance of the corresponding guild, as determined by
their respective phylogenetic trees.

As expected, this is the case for the first iterations, shown in both Figs.%
\ref{parorden1} and \ref{parorden2}. The system is very near the initial
condition, where the contacts have been randomly distributed with no
relation whatsoever to the phylogenetic tree of either guild.

It is interesting to compare such value with the one obtained by using the
observed pattern of contacts of the NCOR system. The values obtained are $%
D^A \simeq D^P\simeq 0.98$ that are indistinguishable from those obtained
for random adjacency matrices.

On the other hand, values significantly lower than 1 as those seen in Fig. %
\ref{parorden1} after many iterations of the SNM, indicate that species that
share the same counterparts are close phylogenetic neighbors.

The ordered pattern that prevails asymptotically for the MAX-CPR case is a
single, perfectly nested pattern~\cite{nosotros2}. After some initial
fluctuations $D^{A,P}$ stabilize very closely to 1. The ordered pattern
turns out to be essentially unique except for a random permutation of
phylogenetic labels and therefore the error bars tend to diminish as the
number of iterations of the SNM increases.

This is not the case for the MIN-CPR in which error bars do not diminish in
the same way. The reason is that there are several possible distinct
asymptotically ordered patterns, all of them corresponding to modular
ecosystems and all having slightly different values of $D^A$ and $D^P$. The
phylogenetic SNM is seen to converge to any of these ordered patterns thus
producing a somewhat larger dispersion in the values of these parametes. In
all cases however, the limiting constant values that are reached greatly
differ from 1 that is many standard deviations away.

\begin{center}
\begin{figure*}[t!]
\includegraphics [width=18 cm]{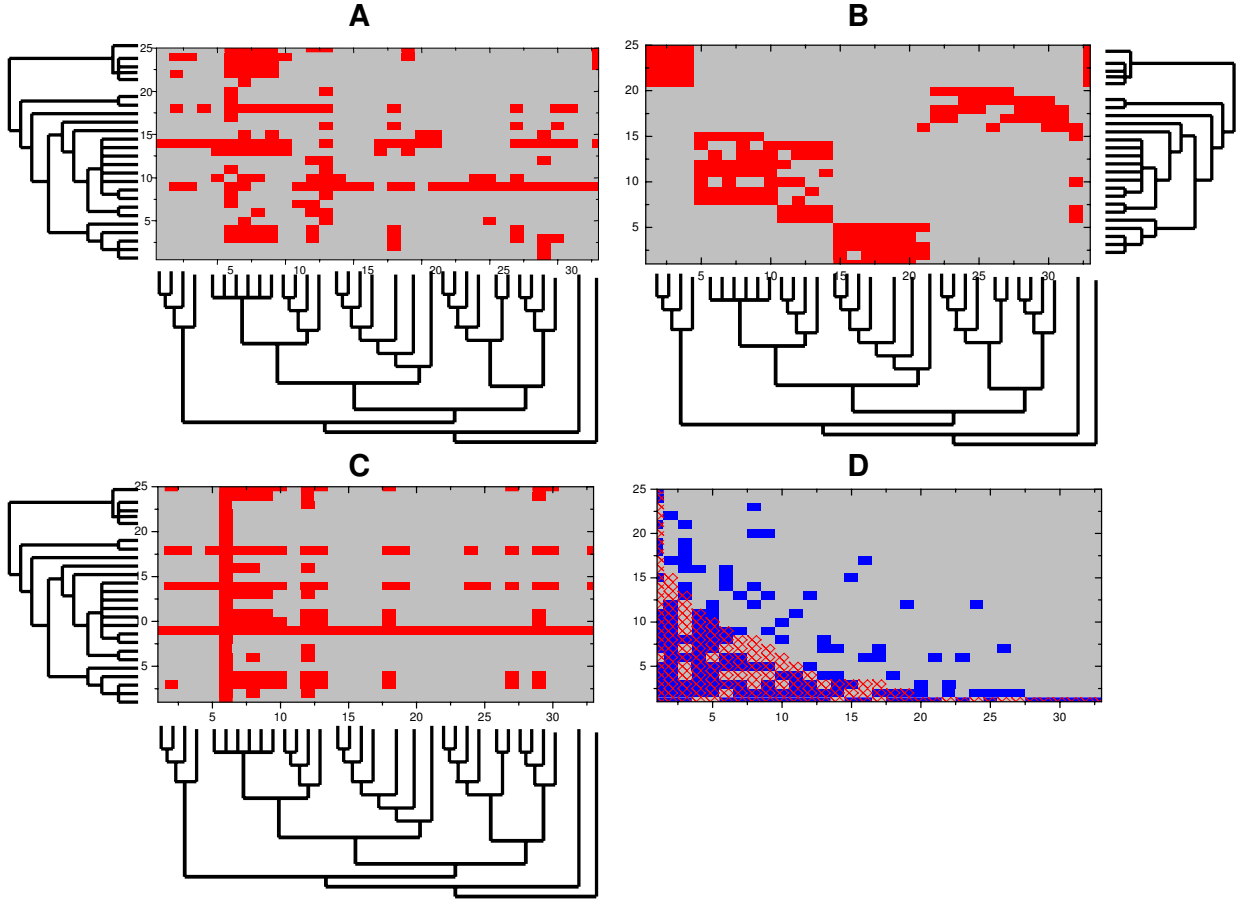} 
\caption{Several adjacency matrices for the study of the NCOR ecosystem.
Panel(A): the empirical contact pattern with species ordered according to
the phylogenetic tree (shown along both margins of the matrix); panel(B):
contact pattern produced by the SNM after 80,000 iterations using as input
the empirical matrix shown in panel (A) and the MIN-CPR alternative, species
in the same order as in panel (A); panel (C): contact pattern produced by
the SNM after 5000 iterations, using MAX-CPR, species in the same order as
in panel (A); panel (D) same contact pattern as panel (A) (dark pixels) and
(C) (hatched pixels) but species are ordered by their degree. Dark pixels
correspond to observed contacts (panel (A)) while slanted pattern
corresponds to the theoretical results (panel (C)). Notice that while in
panel (C) there are generalists ans specialists, they don't show up in panel
(B).}
\label{adyacencia}
\end{figure*}
\end{center}

The SNM algorithm can also be used to test if some given pattern of contacts
is compatible with a CPR involving some kind of phylogenetic dependence.
This can be made through the study of its \textit{stability}. The asymptotic
contact pattern can give a clue of what kind of pattern one should expect to
find for each prevailing CPR. In Fig.\ref{adyacencia} we show the adjacency
matrix of the SNM system as obtained after a great number of iterations
(panels (B) and (C)). These are the asymptotic contact patterns obtained
when applying the MIN-CPR and the MAX-CPR respectively. Since we are
performing a stability analysis we have always used as initial conditions
the empirically observed contact pattern that is shown in Panel (A). These
matrices correspond to configurations in which the parameters $D^{A,P}$ have
reached an almost stationary value and are therefore nearly optimal in the
sense explained in the preceding section. In the same figure are shown the
phylogenetic trees taken from Ref. \cite{Rezende} of plants and animals to
guide the eye. While in panels (A), (B) and (C) the rows and columns of the
adjacency matrix have been ordered as in the tips of the corresponding \
phylogenetic trees, in panel (D) the species of both guilds have been
instead ordered according to their degree with the purpose of rendering more
evident the nested pattern of contacts.

The MIN-CPR corresponds to a rule in which the search of contacts is
dominated by phylogenetic proximity. To better understand the emerging
contact pattern shown in Panel (B), one has to bear in mind that both
animals and plants are considered on equal footing. This gives rise to an
adjacency matrix that breaks into disconnected blocks in which
phylogenetically close species of one guild interact with a similar group of
the other guild. This is the opposite of a nested scheme since the species
tend to specialize its contacts. By the same token, generalists are ruled
out of the system. The contact pattern of the NCOR system used as an initial
condition, becomes therefore severely disturbed putting in evidence that it
is unstable under the presence of the MIN-CPR in which phylogenetic
proximity is the dominant rule.

A similar analysis for the MAX-CPR situation shows an opposite behavior. The
SNM causes no drastic reorderings, reinforcing instead the presence of
generalists and keeping the matrix mildly changed (see panel C). The
observed adjacency matrix must therefore be considered stable under such
CPR. This run of the SNM also provides additional information. The NCOR
system hosts a group of animals that are phylogenetically close and that are
all farely good generalists (e.g. the \textit{turdus} group). Such
correlation between degree and phylogenetic proximity is not destroyed by
the perturbations introduced by the SNM, \textit{if the prevailing CPR is of
the MAX type} (compare panels (A) and (C)).

An additional effect of the iteration of the MAX-CPR rule is that it leads
to an asymptotically stable contact pattern that is almost perfectly nested.
As mentioned before, in panel (D) we show the adjacency matrices of panels
(A) and (C) in which species have been reordered according to increasing
degree; in this way we can compare the empirically observed nested structure
of the NCOR system with an asymptotically nearly perfect nested pattern
produced by the SNM using the MAX-CPR. Since the observed NCOR system is
considerably nested, the effects of the SNM are not drastic.

The occurrence of phylogenetically close species with similar degrees and
the occurrence of a nested pattern should therefore be considered as
independent from each other. When speaking of a \textit{cause} for
nestedness one should expect an element that is present in \textit{all} the
observed systems, with perhaps minor variations in few individual cases. We
have found here that the MAX-CPR strategy always converges to a nested
pattern.

However this alternative combines phylogenetic effects with those just
mentioned due to the number of contacts. This is because contacts are
relocated according to a larger sum $S_{1}$ or $S_{0}$; and this occurs not
only by involving species that are more phylogenetically distant but also by
involving a \textit{greater number} of counterparts as reported in \cite%
{nosotros1}. In order to separate both effects we have made an alternative
test dividing both sums by the number of counterparts that are found in both
rows. In this way contacts are placed in those positions that correspond to
a greater \textit{average} phylogenetic distance. This test completely
separates phylogenetic influence from any other. The results of these
calculations point in the direction that a stationary stable contact pattern
is never reached thus confirming what was said in Ref.\cite{nosotros1}
namely that the chief effect leading to a nested configuration is that all
species tend to place their contacts with already crowded counterparts.
Since on the other hand nestedness is not destroyed by the self organization
process, phylogenetic proximity has therefore to be considered \textit{%
compatible} with a rule inducing nestedness but is far from being a cause of
it.

The above arguments have a greater reach because they hold for \textit{any}
distance matrix or similarity measure. We have made a separate test by
checking the stationary contact patterns that are obtained by introducing
alternative phylogenetic trees. We have considered trees that display a
uniform branching rate and another in which all species successively stem
from a single branch. Notice that in the case of a star phylogeny the self
organization algorithm becomes identical to that of Ref.\cite{nosotros1}. We
have found that in all cases the MIN-CPR strategy leads to modular
ecosystems in which contacts gather in nearly disconected groups. As long as
it is imposed that contacts have to take place between species that are
close to each other, according to some criterion, some kind of
specialization is favored and nestdness turns out to be hampered and not
favored. The opposite is also true: whenever contacts take place with a
greater variety of counterparts, nestedness can be expected to occur.

\section{Conclusions}

The general conclusion that stems from the SNM is that an interaction
between species that exclusively prefers phylogenetic proximity can never
give rise to a nested contact pattern. Nestedness turns out to be unstable
in the presence of such interaction rule. This can be seen by noting that
such interaction mechanism relies in the generalized occurrence of species
that are specialists and ruling out generalists that are an indispensable
ingredient of a nested organization. A CPR governed by phylogenetic
proximity for \textit{both mutualists guilds} tends to destroy a nested
pattern of contacts giving rise to adjacency matrices with a clear tendency
to break down into separate, nearly independent components in which groups
of phylogenetically close neigbors of both guilds hold contacts among each
other but not with the rest of the species of the ecological system.

We have also shown that an alternative interaction pattern dominated by 
\textit{phylogenetic diversity} is instead a much better approach to
describe real situations. This interaction mechanism is one in which species
hold contacts with the greatest possible diversity of mutualist counterparts
that are already visited by a greater number of of species. This rule is
fully consistent with the ones tested in Refs.\cite{nosotros1}, \cite%
{nosotros2}. Highly realistic degree distribution functions and contact
patterns are produced by the SNM with only that assumption.

If a group of phylogenetically close species happen to have similar contact
patterns, this turns out to be stable under such maximal diversity
interaction rule. In fact, a set of phylogenetically close species that also
are farely good generalists remains stable under the organization algorithm
of the SNM. We thus found, in agreement with Ref.\cite{animal}, that
phylogenetic affinity is \textit{compatible} with a nested pattern of
contacts, therefore explaining statistically significant correlations
between degree distributions and phylogenetic proximity. However the results
of the SNM place serious doubts to consider that such correlations are a
sign of causation. The few circumstances in which they have being found to
be statistically significant \cite{Rezende}, \cite{comment}, point into the
direction of considering that these are largely accidental.

We have also tested the present model in other mutualistic systems; the
results are identical to those of the NCOR system and therefore we omit them
here. A dominant cause of the generalized nestedness found in mutualistic
ecosystems perhaps lies on the simple fact that species that we observe in
real systems today are those that tend to put the least possible
restrictions on their mutualist counterparts.

\section*{Acknowledgments}

The authors wish to acknowledge helpful discussions and criticism from D.
Medan and M. Devoto to the original version of the paper to make it more
clear.

\section*{Appendix: the ultrametric distance matrix}

If a tree-like diagram is provided it is also possible to extract from it a
square matrix $d(k,k^{\prime })$ of all ultrametric distances between any
two living species $k$ and $k^{\prime }$. One biologically plausible way to
define such distance is to extract it from their evolutionary history. This
amounts to consider that two species are \textquotedblleft separated by a
distance\textquotedblright\ that is measured by the time elapsed since they
were differentiated in the course of evolution. This distance statisfies a
modified triangular inequality: for any three species $k,k^{\prime }$ and $%
k" $, $d(k,k")\leq \mbox{Max}\{d(k,k^{\prime });d(k^{\prime },k")\}$ that
should be compared to the \textit{metric} triangular inequality $d(k,k")\leq
(d(k,k^{\prime })+d(k^{\prime },k"))$

The evolutionary time can be represented by the length of the branches of
the tree. The ultrametric distance between any two species $k$ and $%
k^{\prime }$ is therefore given by the total length of the branches that
have to be climbed starting either from $k$ or $k^{\prime }$ until a common
ancestor is found. It is of course clear that since the tree-like diagram is
compatible with the analysis of the phylogenetic signal, the ultrametric
distance extracted from the same phylogenetic tree is also compatible with
those stemming from statistical analyses. Resemblances and differences
measured by this ultrametric distance could be considered to involve a
compound effect of all the traits that where considered in the analyses that
lead to the phylogenetic tree.

To obtain the ultrametric distance, a length should be ascribed to the
branches of the phylogenetic tree. Since the true lengths of these segments
are in general unknown some model assumption has to be made. In comparative
studies there is not a universally accepted criterion \cite{Blomberg}
concerning this point and in many analyses all the lengths of the branch
segments are set equal to a constant value since this may be considered to
make fewer assumptions about the data.

In order to get a square matrix with all the ultrametric distances one has
to provide a uniform time scale for all branches, i.e., to provide a time
order for all the branching points of the phylogenetic tree. The most
parsimonious way of doing this is by defining that all branches that stem
from a common ancestor and reach the tips of the tree must have the same
length, counting lengths by starting from the tips and climbing upwards.
This assumption is consistent with the constancy of an evolutionary clock%
\textit{\ \ \cite{kimura}.}

With this procedure the distance matrix can directly be read from the
topology of the phylogenetic tree. We exemplify this procedure in Fig.\ref%
{arbol1}. We define that the two branches that lead to the species labelled
(4) and (5) having a common ancestor in node (A) have a length equal to 1.
By the same rule, the branch starting at species (3) that has a common
ancestor with (4) and (5) in the branchpoint (B) has a length equal to 2.
Moreover, the total length of the branches that have to be climbed starting
from (1) or (2) to reach a common ancestor to all species in (C) must then
have a total length of 3. In all these cases the lengths are defined except
for an overall multiplicative scale factor. This ambiguity is however not
relevant for the present analysis.

\begin{center}
\begin{figure}[tbp]
\includegraphics[width=8 cm]{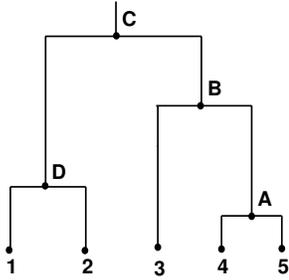} \vspace{-1.5cm}
\caption{An example of a simple phylogenetic tree is shown. The matrix of
ultrametric distances is: $d(1,2)=d(4,5)=1$ $%
d(1,3)=d(1,4)=d(1,5)=d(2,3)=d(2,4)=d(2,5)=3$ $d(3,4)=d(3,5)=2$}
\label{arbol1}
\end{figure}
\end{center}

\end{document}